# IDM-Follower: A Model-Informed Deep Learning Method for Long-Sequence Car-Following Trajectory Prediction

**Yilin Wang**
ORCID: 0000-0002-6002-0535
Lyles School of Civil Engineering, Purdue University
550 Stadium Mall Drive, West Lafayette, IN 47907
Email: wang4517@purdue.edu

**Yiheng Feng, Corresponding Author**
ORCID:0000-0001-5656-3222
Lyles School of Civil Engineering, Purdue University
550 Stadium Mall Drive, West Lafayette, IN 47907
Email: feng333@purdue.edu

**ABSTRACT**
Model-based and learning-based methods are two major types of methodologies to model car following behaviors. Model-based methods describe the car-following behaviors with explicit mathematical equations, while learning-based methods focus on getting a mapping between inputs and outputs. Both types of methods have advantages and weaknesses. Meanwhile, most car-following models are generative and only consider the inputs of the speed, position, and acceleration of the last time step. To address these issues, this study proposes a novel framework called IDM-Follower that can generate a sequence of following vehicle's trajectory by a recurrent autoencoder informed by a physical car-following model, the Intelligent Driving Model (IDM). We implement a novel structure with two independent encoders and a self-attention decoder that could sequentially predict the following trajectories. A loss function considering the discrepancies between predictions and labeled data integrated with discrepancies from model-based predictions is implemented to update the neural network parameters. Numerical experiments with multiple settings on simulation and NGSIM datasets show that the IDM-Follower can improve the prediction performance compared to the model-based or learning-based methods alone. Analysis on different noise levels also shows good robustness of the model.

## INTRODUCTION
Car-following is the most common driving behavior which attracts tremendous research interest in the transportation community for decades. Numerous car-following models have been developed to describe human driving behaviors. Typical car-following models output the location, speed, or acceleration of the following vehicle in the next time step with the considerations of both the leading vehicle and following vehicle's status such as position and speed, and driver's characteristics such as reaction time at the current time step. Based on different modeling approaches, most existing studies can be classified into two categories: model-based and learning-based. Model-based approaches describe the car-following behaviors with explicit mathematical equations while learning-based approaches usually apply machine learning techniques to map the input and output relations by training neural networks with observed trajectory data.

Model-based and learning-based models have strengths and weaknesses in predicting car-following behaviors. Model-based models are more robust and consistent with prevalent driving behaviors. The explicit model form with a small set of parameters makes model-based car-following models easy to be implemented and calibrated. In addition, these models can be easily integrated with other applications. For example, Newell's first-order car following model (*1*) is widely used in other traffic models to calculate vehicle trajectories (e.g., as constraints in the optimization problem). However, model-based methods don't consider heterogeneity in driving behaviors, and thus may perform poorly in cases with atypical following behavior and/or noisy data. On the other hand, learning-based methods can handle heterogeneous driving behaviors with large parameter sets and complex model structures. Learning-based methods usually require a large amount of data for training, validation, and testing. With more and more sensors being installed on both smart vehicles and infrastructure side, acquiring enough training data may not be an issue in the future. But lack of interpretability is one major shortcoming for learning-based methods since the mapping relations within the neural network is very difficult to explain. As a result, we usually consider these models as "black boxes". Moreover, when the training data set is noisy or far away from the ground truth, learning-based methods perform poorly. To leverage the advantages of both modeling approaches, physics or model informed machine learning methods have been proposed in the last few years. This new type of model integrates physical models (e.g., partial differential equations) with machine learning models and outperforms both model-based and learning-based methods (*2–4*).

In this paper, we propose a model-informed deep learning method, called IDM-Follower, to predict long sequence car-following behaviors. Particularly, we integrate the intelligent driving model (IDM) with the sequence-to-sequence recurrent autoencoder to predict the following vehicle trajectory. The autoencoder is constructed by two encoders and one self-attention decoder. It extracts features from multiple state sequences and then decodes with hidden layers to generate a sequence output. We design a hybrid loss function that integrates the calibrated car-following model when training the autoencoder. We test the proposed method on a simulation dataset and NGSIM dataset with multiple GPS noise levels. Results show that IDM-Follower has better performance than the learning-based baseline. Additionally, it shows promising performance in terms of robustness against different levels of noise.

To the best of our knowledge, the contributions of this paper are as follows:
1. We propose a novel framework that integrates sequence-to-sequence deep learning based prediction model with a traditional car-following model.
2. By utilizing the self-attention mechanism, the proposed model can address long-term historical dependency and handle the long-sequence trajectory prediction problem.

3. The proposed model is robust against GPS noises, which is applicable for many transportation applications.

The rest of the paper is organized as follows: Section 2 reviews related studies on car-following modeling and prediction. Section 3 describes details of the proposed model-informed deep learning framework including architecture, neural network structure and loss function design. Section 4 presents two numerical experiments based on simulation data and NGSIM data. Section 5 summarizes the paper with future research directions.

**LITERATURE REVIEW**
A car following model is usually described as a parameterized mathematical mapping between the "state" and "action". The state (defined as $S$) is a time-indexed set including the kinematic variables that describe the movement status of the leading and following vehicles such as position and relative speed. The action (defined as $A$) is another time-indexed set that records the following vehicle accelerations. A car following model usually generates actions for the next time step based on the state variables of the current time step parameterized by $\theta$, expressed as below:
$$f_\theta: S \to A$$
The goal of building a car-following model is to: 1) define a proper mapping function $f$ and 2) find an optimal set of parameters $\theta$, that can describe real-world driving behaviors.

In the middle of the 1950s, Gazis and Herman(*5*) developed the first car-following model (GHR model) with researchers from General Motors. The GHR model assumes that the following vehicle decides its acceleration in a stimulus-response of the relative speed between the leading and following vehicles. After that, multiple notable works including Gipps model (*6*), Wiedemann model (*7*), and IDM (*8*) have been developed and implemented for a wide range of applications such traffic state estimation, traffic signal optimization and vehicle trajectory planning(*9–11*).

With the rapid development on machine learning techniques, as well as the availability of high-fidelity traffic data, researchers began to model the car-following behaviors using learning-based methods, which can be classified into the following three types:

1) Unsupervised learning model: He et al. (*12*) proposed a nonparametric car-following model with k-nearest neighbors. The model outputs the most likely driving behavior under the current circumstance by calculating the average of the most similar cases. The model requires neither calibration nor the assumption of the fundamental diagram that describes the driving behaviors. Papathanasopoulou and Antoniou (*13*) developed a nonparametric learning-based car-following model named the Loess model based on locally weighted regression.

2) Fully connected neural network can predict the vehicle's future actions based on multiple inputs. Studies using fully connected neural networks for car-following behavior prediction can be traced back to 2007. Sakda et al. (*14*) built a neural agent car-following model to map the perceptions to vehicle acceleration. Jia et al. introduced a four-layer neural network (one input layer – two hidden layers – one output layer). The model takes relative speed, follower speed, maximum desired speed, and the gap between two vehicles as inputs to predict the follower's next-step acceleration.

3) Recurrent neural networks (RNN): RNN structures enable complicated feature extraction from the driving data. Zhou et al. (*15*) proposed a vanilla RNN to model drivers' car-following behavior, with a focus on predicting traffic oscillations. Wang et al. (*16*) used a variant of RNN called Gated Recurrent Unit (GRU) (*17*) to model car-following behaviors.

Furthermore, recent studies began to utilize sequence-to-sequence model to predict the car following behavior. Ma and Qu (*18*) proposed a sequence-to-sequence car following prediction model using another variant of RNN called Long Short-Term Memory (LSTM) network. Zhu et al. (*19*) proposed a long-sequence car-following trajectory prediction model with the consideration of historical context. They proposed a transformer framework that could learn the car-following behaviors from the first 4 seconds and predict the rest of the 12 seconds with a transformed decoder.

In the past few years, researchers began to integrate physical models with machine learning models in the transportation domain. Yang et al. (*3*) proposed a stochastic physics regularized Gaussian process (PRGP) model for the macroscopic traffic state estimation. Taking Gaussian process as the learning model, this study implemented a mixed structure of learning framework along with a new loss function including both physical and learning-based losses. The lost function not only calculates the distance between the predictions and ground truth but also the loss between the prediction results and the physical model. Similarly, Shi et al (*4*), and Huang et al(*2*) proposed physics-informed deep learning (PIDL) frameworks that integrate neural networks with the Lighthill Whitham-Richards(LWR) model (*20*) to construct fundamental diagrams of a highway segment. Similar lost functions are designed that combine data and physical losses. The abovementioned studies integrated physical models to regulate the traditional machine learning model for macroscopic traffic state estimation. A recent study (*7*) tries to apply the same framework to integrate neural networks with car-following models. Two representative car-following models, namely IDM and optimal velocity model (OVM) are integrated with a neural network to predict the acceleration of the following vehicle based on the velocity and gap between the leading and following vehicles. Moreover, they developed a novel structure that could simultaneously calibrate the car-following model parameters as well as predict the following vehicle's acceleration. However, this framework can only predict following vehicle accelerations step by step, rather than an entire sequence simultaneously. Inspired by previous studies with the consideration of their drawbacks, this paper proposes a deep learning-based sequence-to-sequence vehicle trajectory prediction method integrated with the IDM car-following model, named IDM-Follower.

**METHODOLOGY**
In this section, we introduce the framework of the IDM-Follower with the problem statement, model architecture, and model implementation.

**Problem Statement**
Define a trajectory pair $z^{(i)} = \{X^{(i)}, Y^{(i)}\}$ with $i$ represents the trajectory index. $X^{(i)}$ and $Y^{(i)}$ separately represent the trajectory for the leading and following vehicles. The leading vehicle's trajectory consists of two time series vectors that represent position and velocity while the following vehicle's trajectory includes only the position vector. These vectors can be represented as a function of time $t$ as shown below:

$$X^{(i)} = \{s_X^{(i)}, v_X^{(i)}\}, Y^{(i)} = \{s_Y^{(i)}\} \; for \; i = 1, \ldots, N$$
$$s_X^{(i)} = \{s_X^{(i)}(t), t = 1, \ldots, T\}, v_X^{(i)} = \{v_X^{(i)}(t), t = 1, \ldots, T\}$$

Where $N$ is the set of the trajectory pairs and $T$ is the length of the trajectories, assuming all trajectories have the same length.

Further, we define a learning-based car follow model parameterized by a hyperparameter vector $\theta$ as $F_\theta(X^{(i)}|\theta)$. The output of this function is the following vehicle's trajectory $\tilde{Y}^{(i)}$ based on the leading vehicle's trajectory $X^{(i)}$. The mapping relation can be expressed as $s_X^{(i)} \in X^{(i)} \rightarrow \tilde{s}_{Y,\theta}^{(i)} \in \tilde{Y}^{(i)}, \forall \, i = 1, \ldots, N$. Note that we denote the predicted quantities with a tilde, to distinguish them from observed values. Similarly, we define a model-based mapping $s_X^{(i)} \in X^{(i)} \rightarrow \tilde{s}_{Y,\lambda}^{(i)} \in \tilde{Y}^{(i)}, \forall \, i =$ $1, \ldots, N$, that is parameterized by a hyperparameter vector $\lambda$. The $F_\theta$ function is represented by a neural network while the $F_\lambda$ function is a set of physics-based equations. The IDM-Follower, a model-informed car following trajectory predictor can be defined as $F_{\theta,\lambda}(X^{(i)}|\theta,\lambda): s_X^{(i)} \in X^{(i)} \to \tilde{s}_Y^{(i)} \in \tilde{Y}^{(i)}, \forall i = 1, \ldots, N$. The goal is to train an optimal parameter set $\theta^*$ for the neural network with regularization from the physical car-following model. In our work, the optimal physical model parameter set $\lambda^*$ is assumed to be given, which can be calibrated from the observed data.

The model-informed deep learning framework consists of a model-based component and a learning-based component as shown in Equations 1 and 2. The model component is encoded as the IDM, while the learning component is a sequence-to-sequence deep neural network. The primary output for the IDM is an acceleration vector for each time step. In our design, the deep learning model outputs a position vector that represents the following trajectory. Since the output of the IDM is an acceleration vector, to be consistent with the deep neural network output, the acceleration vector from IDM is converted to a position vector in Equation 3.

Learning-based component:
$$\tilde{s}_{Y,\theta}^{(i)} = F_\theta(X^{(i)}|\theta) \ \forall i = 1, \ldots, N \quad (1)$$

Model-based component:
$$F_\lambda(X^{(i)}, s_Y^{(i)}|\lambda) = \tilde{a}_Y^{(i)} \ \forall i = 1, \ldots, N \quad (2)$$
$$\tilde{s}_{Y,\lambda}^{(i)} = \iint \tilde{a}_Y^{(i)} \, dt \quad (3)$$

Although sequence-to-sequence neural networks allow different sizes of dimensions for input and output data, in this work, we assume that the prediction horizon (size of the output sequence) is the same as the input sequence. Given the above definitions, the car-following trajectory prediction can be formulated as the following optimization problem in equation 4.

$$\min_\theta \sum_{i=1}^N \left\| \tilde{s}_Y^{(i)} - s_Y^{(i)} \right\|_2 \quad (4)$$
$$\text{s.t } \tilde{s}_Y^{(i)} = F_{\theta,\lambda^*}(X^{(i)}|\theta,\lambda^*), \forall i = 1, \ldots, N$$

**Model-informed Trajectory Prediction Architecture**

This section provides an overview of the model-informed trajectory predictor. Figure 1 shows the details of workflow. The state sequence of the leading vehicle is fed into a neural network, which consists of neurons associated with activation functions parameterized by $\theta$. The objective is to train the optimal parameter set $\theta^*$ using observed data. On the other hand, the state sequence of the leading vehicle is fed into a car following model to calculate the following vehicle's trajectory. The parameter set of the car-following model $\lambda^*$ is calibrated in advance, so that no training is needed. Both models calculate the training loss. One loss represents the differences between the predicted trajectories from the neural network and the observed trajectory. The other loss represents the differences between the predicted trajectories from the physical car-following model and the predicted trajectories from the neural network. Then, the two losses are combined as the total loss via a linear combination with a weight parameter $\mu$. Finally, the gradient of the combined loss function is calculated and used to update the parameters of the neural network through backward propagation. Compared with the traditional learning-based framework, the model-informed framework takes advantage of the physical car-following model to regularize the search space as well as consider the complex data patterns from the observations. This framework is generic and can be applied to any sequence-to-sequence deep learning model and car following.

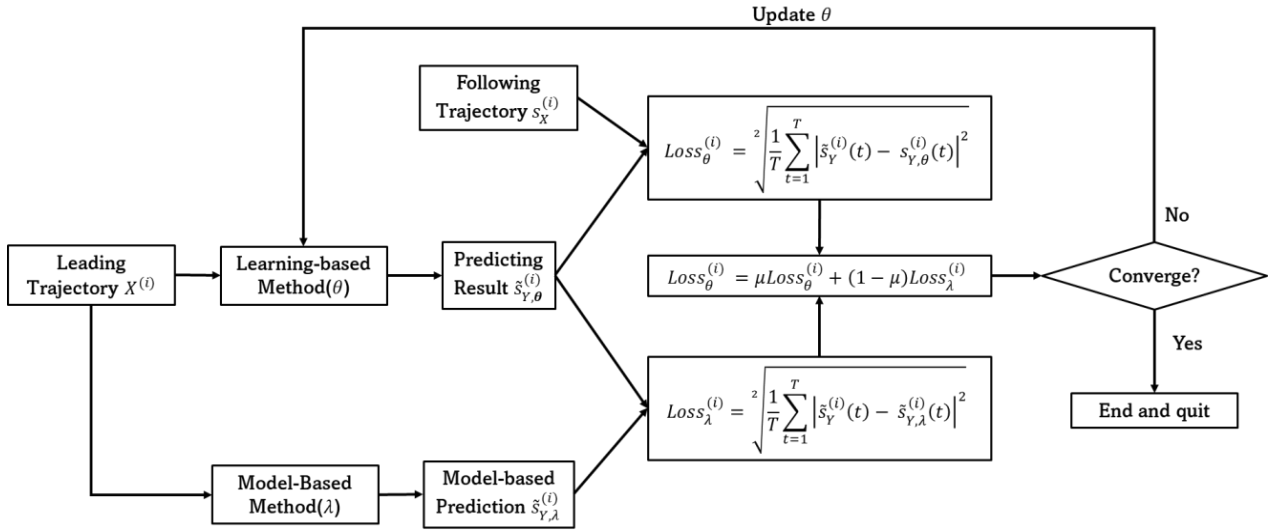

**Figure 1:** Workflow of IDM-Follower

## Neural Network Structure

Sequence-to-sequence models generate a patterned sequence from a given sequence. It can extract and model the relationship between two sequences into a high-dimension parameter set in a neural network architecture. Autoencoder (AE), a symmetric encoder-decoder model, is able to extract the feature patterns from the input data into an encoder. Then the decoder reconstructs the data using the encoder's output as the initial input. Due to its capability of mapping a complex relationship between similar types of data (i.e., clean, and noisy images, two sentences from different languages), AE has been widely used in noisy data processing and machine translation (*21*, *22*). Due to its recursive architectures, typical recurrent neural networks factor computing process with the symbolic positions from the input and output sequence, making it possible to process sequential data. Generating a sequence of hidden states $h_t$ as the function of the previous hidden state $h_{t-1}$ and the inputs at *t*, the inherently sequential nature precludes parallelization within training examples. However, the model performance becomes problematic for long sequence data as the model memory constraints limit batching across examples. This issue becomes obvious in recurrent autoencoder (RAE) since the symmetric decoder-encoder structure greatly adds up to the model complexity. Vehicle trajectories, however, vehicle trajectory is a typical long sequential type of data. To address this challenge, we apply self-attention, a mechanism that would relate different positions of a single sequence, to address the memory limitations and error accumulations for long sequence data. It has been successfully applied in various tasks including reading comprehension (*23*) and machine translation (*24*).

The encoder will encode the input state (e.g., leading vehicle trajectory) to a hidden layer, where the features of the sequence will be extracted and restored, after going through all the LSTM layers in the encoder. Different from conventional autoencoder with one encoder, this paper designs two disjoint encoders that separately encode the position sequence and the velocity sequence, since most of the car following models take both the leading vehicle's position and speed into account. With independent encoders on position and speed, the decoder takes two hidden layers from the encoders to reconstruct the following vehicle's trajectory.

The decoder is designed with a self-attention mechanism. Three attributes including query, value, and key are required to process the attention layer. The attention layer then uses this input to generate a context vector which is used as the input of the following decoder. Then, the hidden state vector will be decoded from the position sequence decoder with another LSTM layer to generate the query. The value will be applied by the hidden layer of the position encoder and the key is the encoding output of the velocity encoder. In summary, the structure of the neural network is shown in Figure 2, and its parameters are listed below.

*Encoder:*
LSTM (input size=1, hidden size=128, layers=2)
key network: Linear (256, 128)
value network: Linear (256, 128)
*Decoder:*
LSTM (input size=128, hidden size=128, layers=2)
Attention (key size = 128, value size = 128)
output layer: Linear (256, 1)

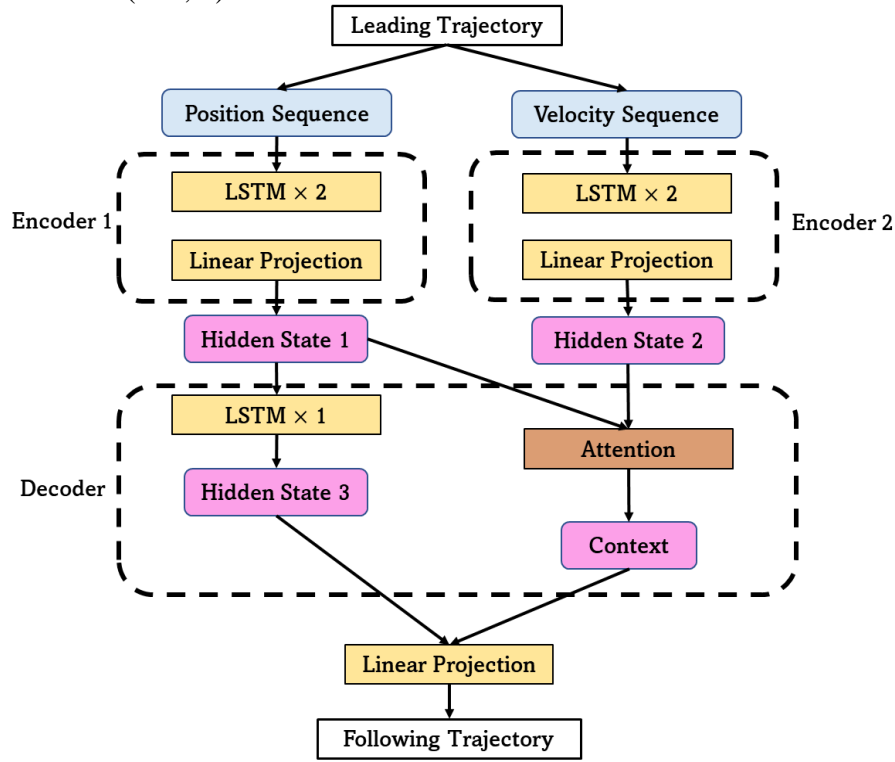

**Figure 2:** Structure of the self-attention autoencoder

**Loss Function**
The neural network updates the parameter set based on the gradient of the loss function. The loss function is composed of two parts. The objective of learning-based loss, defined as $L_{data}: \mathbb{R}^{1 \times n} \to \mathbb{R}^{1 \times n}$, is to minimize the discrepancy between the observed trajectory and neural network prediction. The model-based loss, defined as $L_{model}: \mathbb{R}^{1 \times n} \to \mathbb{R}^{1 \times n}$, is designed to calculate the discrepancy between the physical model prediction and the neural network prediction. Specifically, the data discrepancy for both losses is evaluated by the rooted mean square error (RMSE). The total loss is defined as the weighted sum of the two parts, as shown in equation 5.

$$Loss_\theta^{(i)} = G\left(L_{data}^{(i)}, L_{model}^{(i)}\right) = \mu L_{data}^{(i)} + (1-\mu)\, L_{model}^{(i)}$$

$$= \sqrt{\mu^2 \frac{1}{T}\sum_{t=1}^{T} \left|\tilde{s}_Y^{(i)}(t) - s_{Y,\theta}^{(i)}(t)\right|^2} + \sqrt{(1-\mu)^2 \frac{1}{T}\sum_{t=1}^{T} \left|\tilde{s}_Y^{(i)}(t) - \tilde{s}_{Y,\lambda}^{(i)}(t)\right|^2}$$

$$= \mu^2 \sqrt{\frac{1}{T}\sum_{t=1}^{T} \left| f_\theta(X^{(i)}|\theta) - s_Y^{(i)}(t) \right|^2} + (1-\mu)^2 \sqrt{\frac{1}{T}\sum_{t=1}^{T} \left| f_\theta(X^{(i)}|\theta) - \iint_1^t f_{\lambda^*}\left(X^{(i)}, s_Y^{(i)} \middle| \lambda^*\right) dt \right|^2} \quad (5)$$

## NUMERICAL EXPERIMENTS

In this section, we present two numerical experiments that evaluate the performance of the model-informed deep learning framework with both simulation and real-world datasets. The simulation data are generated by SUMO and the real-world car-following trajectories are extracted from the NGSIM Lankershim Blvd dataset. GPS noises are added to the original trajectories to mimic the localization error.

### GPS Noise Generation

GPS errors come from various sources including signal arrival, ionospheric effects, ephemeris effects, and multipath distortions according to Farrell and Barth (*25*). In downtown areas with tall buildings, the GPS errors may cause a completely offset to another street from the real location. It is important that the proposed method is robust against GPS noises. We adopt the GPS error model developed by Feng (*26*) to add offsets to both simulation and NGSIM data. In the model, latitude and longitude GPS points are independently calibrated with two univariate autoregressive integrated moving average (ARIMA) models respectively. Since this study mainly considers one-dimensional driving behavior, only latitude GPS error is added. Table 1 shows three sets of parameters for generating GPS errors from the ARIMA model. The small noise level represents the original GPS error calibrated from the field data while the other two levels are intended to simulate higher noise levels to further test the robustness of the proposed method. SD represents the intensity of the fluctuations around the ground truth positions while the intercept determines the displacement error.

**Table 1:** ARIMA model parameters for different noise level

| Noise level | GPS MAE | AR1 | AR2 | MA1 | MA2 | SD | Intercept |
|---|---|---|---|---|---|---|---|
| Small | 1.79 | -0.9548 | -0.3673 | 0.9188 | 0.3163 | 1.16074 | 1.7923 |
| Middle | 5.63 | -0.9548 | -0.3673 | 0.9188 | 0.3163 | 1.16074 | 5.3769 |
| Big | 10.48 | -0.9548 | -0.3673 | 0.9188 | 0.3163 | 1.16074 | 10.7538 |

### Testing on Simulation Dataset

*Data collection and experiment settings*

The simulation dataset is generated from SUMO, a microscopic traffic simulation environment. A one lane road is built with a fixed-time traffic signal at the end to create a bottleneck and generate different driving behaviors including free flow, accelerating, braking, and following with constant speed. Since SUMO uses IDM as its default car following model, we directly retrieve the IDM parameters from SUMO as the optimal parameter of the model-informed part ($\lambda^*$), as shown in the first row of Table 2.

We consider the maximum car-following distance is 50 meters when extracting car-following trajectory pairs. If the gap between two vehicles is greater than 50m, we consider the following vehicle is in the free flow state. Then their trajectories are not included in the dataset. All qualified leading-following trajectory pairs are processed to 8s sequences (e.g., 80 points), which is the prediction horizon. Then different levels of GPS errors are added. The model-informed framework will take the noisy vehicle trajectories as the input data and the noisy following vehicle trajectories as the label for training. A total of 8942 pairs of trajectories are selected to build the simulation dataset. The number of trajectory pairs for training, validation, and testing are 4471, 1788, 2682 with a ratio of 0.5, 0.2, and 0.3 respectively.

**Table 2:** Optimal IDM parameters used for different datasets

| Dataset Name | $v_0$ | $T$ $(s)$ | $s_0$ | $a$ | $b$ | $\delta$ | FDE |
|---|---|---|---|---|---|---|---|
| SUMO Dataset | 16.7 | 1 | 2.5 | 3 | 4.5 | 4 | 2.75 |
| NGSIM Dataset 1(27) | 15.97 | 1.3 | 1.57 | 2.49 | 2.39 | 4 | 3.47 |
| NGSIM Dataset 2(28) | 12.58 | 0.48 | 0.31 | 1.98 | 4.37 | 1.34 | 5.23 |

FDE: Final Displacement Error (m)

The training settings are listed in Table 3 and Figure 3 shows the training loss under the small noise level for both training and validation datasets.

**Table 3:** Training Hyperparameters training in both datasets.

| Hyperparameter Name | SUMO dataset | NGSIM dataset |
|---|---|---|
| Max Epoch | 200 | 300 |
| Optimizer | Adam | Adam |
| Learning Rate | $1.0 \times 10^{-3}$ | $1.0 \times 10^{-3}$ |
| Learning Rate Weight Decay | $1.0 \times 10^{-5}$ | $1.0 \times 10^{-5}$ |
| Bach Size | 64 | 64 |
| Training, Testing and Validating count | (4471, 1788, 2682) | (1730, 494, 741) |

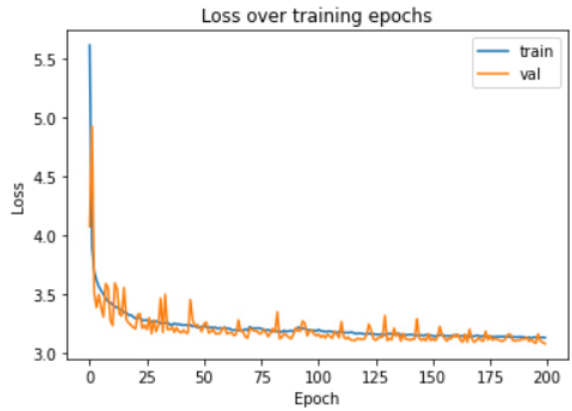

**Figure 3:** Training Loss Plot with SUMO dataset

*Performance Evaluation and Sensitive Analysis*

Figure 4 shows the prediction results from the IDM-Follower. In the figure, the blue curve and orange curve represent the ground truths of leading and following vehicle trajectories. Learning-based Autoencoder represents the baseline results for the pure learning-based method while the Model-based IDM represents the pure model-based method. Although the training is based on the data with noise, the testing errors are calculated based on the prediction results and the ground truth data (without noise). In this figure, the learning-based result shows some unreasonable fluctuations that violate the physical laws such as the trajectory going backward around time point 40s. . The model-based result, on the other hand, is much smoother but the error accumulates over time. The model-informed framework generates its trajectory with fewer fluctuations than the learning-based model while keeping a small error over time.

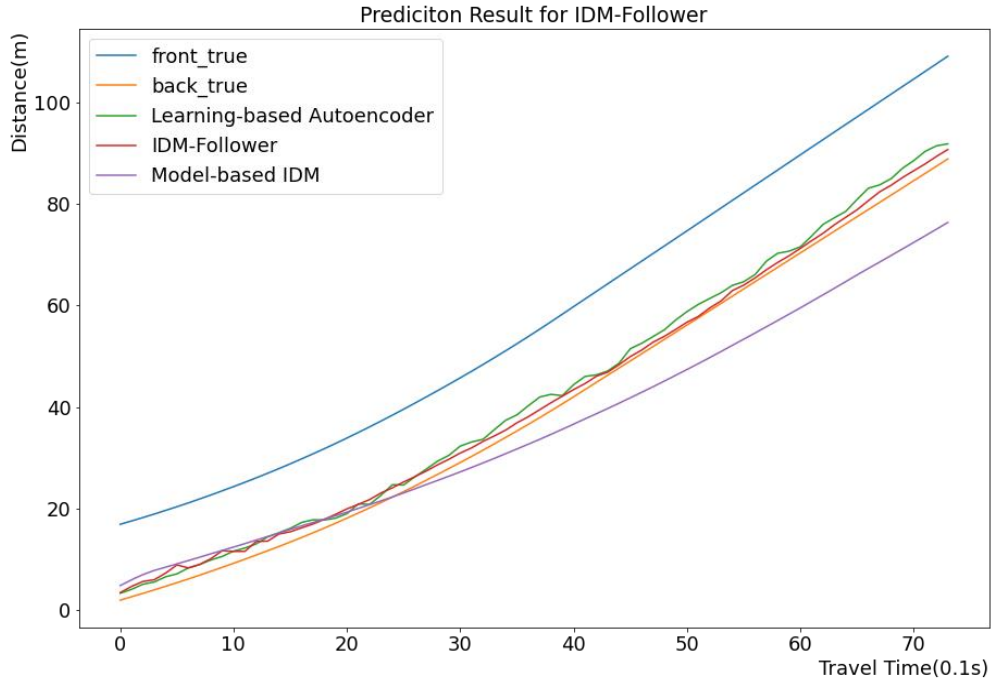

**Figure 4:** Prediction Result for IDM-Follower in SUMO dataset (Small Noise)

To further test the performance of the IDM-follower, we test different weights between the model-based and learning-based components. Rooted Mean Square error (RMSE) (Equation 5) is utilized to evaluate the performance. The RMSE is used for evaluating the sequence similarity, which is commonly used to measure the accuracy of the trajectory prediction.

$$RMSE^{(i)} = \sqrt[2]{\frac{1}{T}\sum_{t=1}^{T}\left|\tilde{s}_{Y,\lambda,\theta}^{(i)}(t) - s_Y^{(i)}(t)\right|^2} \quad (5)$$

In Table 4, cases with different values of $\mu$ that weights between the neural network and IDM model are compared to two baselines, learning-based and model-based IDM. Results show that the IDM-Follower reaches the highest accuracy when 30% IDM is integrated under small noise and 50% under middle noise. The average RMSE is reduced by 2.1% to 26.9% compared to the learning-based baseline in two cases.

**Table 4:** Testing Result from SUMO dataset (RMSE) (m)

| Model Index | $\mu$ | Small noise | Middle noise |
|---|---|---|---|
| Learning-based Method | - | 2.37 | 7.40 |
| 70%Data, 30% IDM | 0.7 | **2.32** | 5.58 |
| 50%Data, 50% IDM | 0.5 | 3.83 | **5.41** |
| 30%Data, 70% IDM | 0.3 | 4.87 | 7.66 |
| Model-based IDM | - | 4.99 | 5.93 |

Different noise levels exert different impacts on learning-based and model-based methods. When the noise level is small, the pure learning-based method performs much better than the IDM model. When the noise level increases, the error of the learning-based method increases dramatically while the performance of the IDM model only gets a little worse. When the impact of the noise becomes bigger under the middle level, the model-based method shows higher robustness. As a result, the optimal weight of integrating the learning-based and model-based components is different in the two cases. Under middle noise level, more weight is put on the model component and the improvement over the learning-based method is also significantly improved (26.9% vs. 2.1%).

Interestingly, adding weights to the model-based method does not always result in an improvement. A negative impact occurs when the IDM weight reaches 50% and 70% under the small and middle noise levels respectively. It is critical to choose a proper weight based on the performance of the learning-based baseline and model-based baseline. Further studies are needed to investigate the optimal contribution ratio under different scenarios.

In summary, tests on SUMO simulation data show that integrating car-following model can improve the prediction accuracy of the learning-based method under different noise levels. The improvement is significant when the model-based method shows better performances than the learning-base method. However, these experiments are conducted in a simulation environment. All the trajectories are generated by the IDM model with the same parameter set of the model-based component. In real-world driving environments, the driving behaviors vary among drivers, which have higher uncertainties and are more difficult to predict for both model-based and learning-based methods. Next, we present the testing results based on the NGSIM dataset.

**Testing on NGSIM Dataset**
*Data Process and Experiment Settings*
We test the proposed framework on the NGSIM Lankershim Boulevard dataset. Bidirectional trajectories of the three to four lanes in an approximately 1600 feet street in the coverage of three signalized intersections are collected. A total of 30 minutes of raw data are available in the dataset. We first apply Kalman Filter(*29*) to clean the noises due to video processing. The output data is considered as ground truth vehicle trajectories. Similar to the simulation dataset, qualified trajectory pairs are selected and processed to 8s sequences. Then GPS noises are added to ground truth trajectories to create the training dataset. In total, 2472 pairs of trajectories are selected. The number of trajectory pairs for training, validation, and testing are 1730, 494, and 741 with a ratio of 0.5, 0.2, and 0.3 respectively. Different from the simulation data that the optimal IDM model parameters can be obtained directly from SUMO, for the NGSIM dataset, the IDM parameters used for the model component need to be calibrated. We adopt the calibration results from an existing study (*27*) shown in the second row of Table 2. We further validate the model parameters by calculating the average final displacement error (FDE) in our processed trajectory pairs. The average FDE is 3.47m, which is an acceptable error.

*Performance Evaluation and Sensitive Analysis*
For the NSGIM dataset, all three noise levels are added to the ground truth data and three models are trained for different noise levels Figure 5 shows the training loss plot for the NGSIM dataset under each noise level. From left to right of Figure 5, the noise level increases.

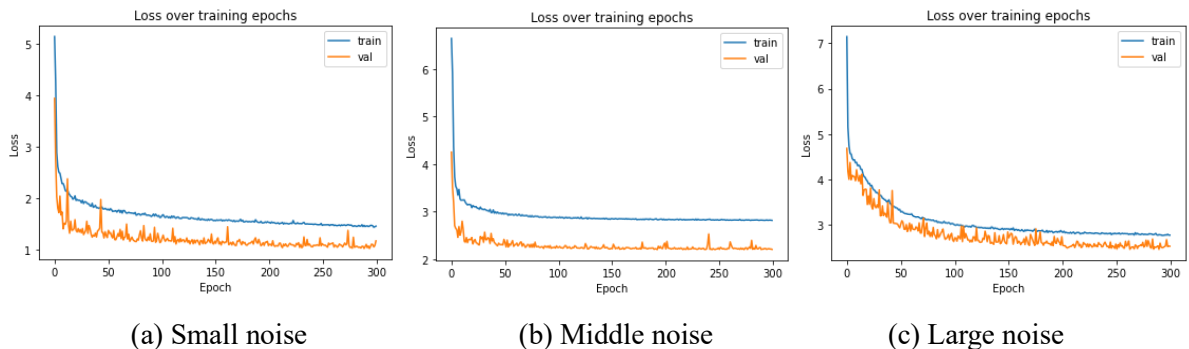

(a) Small noise  (b) Middle noise  (c) Large noise
**Figure 5:** Training Loss Plot with NGSIM dataset under different noise levels

Similarly, RMSE is utilized to evaluate the performance. The results are shown in Table 5. When the noise is small, the learning-based method and 30% IDM achieve the best performance.

However, when the noises get bigger, integration of the model component improves the performance. This is because the model-based method is less sensitive to the increasing GPS noise than the data noise, due to physical constraints embedded in the model. It is noted that under different noise levels, the optimal value for the weight $\mu$ is always 0.7. To further investigate the influence of different parameters from the model component on the model-informed method, we apply another parameter set from an existing study (*24*) as shown in the third row of Table 2.

**Table 5:** Testing Result for NGSIM dataset with multiple noise conditions (RMSE)

| Model Index | $\mu$ | Small noise (m) | Middle noise(m) | Big noise(m) |
| --- | --- | --- | --- | --- |
| Learning-based Method | - | **3.52** | 6.13 | 12.83 |
| 70%Data, 30% IDM | 0.7 | **3.52** | **5.56** | **11.79** |
| 50%Data, 50% IDM | 0.5 | 4.69 | 7.54 | 13.43 |
| 30%Data, 70% IDM | 0.3 | 6.67 | 10.99 | 15.56 |
| Model- based IDM | - | 5.75 | 5.74 | 12.86 |

Table 6 shows a comparison between the performance of IDM-Follower with different IDM parameter sets. Two calibrated parameters from existing studies (Wang, et al 2021 and Yang, et al 2022 ) (*27, 28*) are used for the testing with an average FDE of 3.47m and 5.98m respectively (Table 2). For each noise level, we chose μ=0.7 since 30% integrated IDM shows the highest accuracy for each noise level. Since Wang's parameter set has less FDE than Yang's parameter on our testing dataset, we consider Wang's calibration has better performance. The results show that a less accurate model parameter set also results in worse performance in terms of RMSE in all cases. The results are consistent with (*3*) that an improper physical model may degrade the performance of the model-informed learning framework.

**Table 6:** Testing Results of NGSIM dataset with different IDM parameter sets (RMSE) (m)

| Noise level | Model | $\mu$ | IDM parameter from (23) | IDM parameter from (24) |
| --- | --- | --- | --- | --- |
| Small noise | 70%Data, 30% IDM | 0.7 | **3.52** | 4.04 |
| Middle noise | 70%Data, 30% IDM | 0.7 | **5.56** | 6.03 |
| Big noise | 70%Data, 30% IDM | 0.7 | **11.79** | 13.84 |

## CONCLUSIONS

In this paper, we proposed a supervised model-informed machine learning framework to predict long sequences of following vehicle trajectories with noisy observations. We designed an autoencoder based neural network with an attention layer as the learning component and integrated it with the IDM car-following model as the model component through a customized loss function. Then model-informed neural network was trained with both a simulation and the NGSIM datasets under different GPS noise levels. Experiment results showed that the model-informed framework outperforms either pure learning-based or model-based methods when a proper weight is selected.

To the best of our knowledge, this study is the first one that integrated model-based method into a sequence-to-sequence deep learning prediction model. Our future work includes more comprehensive testing and theoretical analysis on how to choose optimal weights between the model-based component and learning-based component under a given scenario setting.

## ACKNOWLEDGEMENT
This research is supported in part by the U.S. Department of Transportation through Grant #693JJ32150006, Smart Intersections: Paving the Way for a National CAV Deployment The views presented in this paper are those of the authors alone.

**AUTHOR CONTRIBUTION STATEMENT**
The authors confirm contribution to the paper as follows: study conception and design: Yilin Wang, Yiheng Feng; data collection: Yilin Wang; analysis and interpretation of results: Yilin Wang, Yiheng Feng; manuscript preparation: Yilin Wang, Yiheng Feng. All authors reviewed the results and approved the final version of the manuscript.